\documentclass{elsart}
\usepackage{natbib}
\usepackage{graphicx}
\usepackage{amssymb}
\usepackage{amsmath}
\usepackage{bm}

\def\url#1{{\ttfamily\def\/{/\discretionary{}{}{}}#1}}
\def\bibcode#1{}

\newcommand\ba{\begin{eqnarray}}
\newcommand\ea{\end{eqnarray}}
\newcommand\eq{\begin{equation}}
\newcommand\en{\end{equation}}

\begin{document}
\begin{frontmatter}
\title{The Formation Histories of Galaxy Clusters}
\author{J.D. Cohn\thanksref{jdcemail}${}^{a,c,}$}  and 
\author{Martin White\thanksref{mjwemail}${}^{b,c,}$}
\address{${}^a$Space Sciences Laboratory, 
${}^b$Departments of Physics and Astronomy,\\
${}^c$Theoretical Astrophysics Center\\
University of California, Berkeley, CA 94720}
\thanks[jdcemail]{jcohn@astro.berkeley.edu}
\thanks[mjwemail]{mwhite@berkeley.edu}
\begin{abstract}
A sample of hundreds of simulated galaxy clusters is used to study the
statistical properties of galaxy cluster formation.  Individual assembly
histories are discussed, the degree of virialization is demonstrated and
various commonly used formation times are measured and inter-compared.
In addition, the fraction of clusters which have ``recently'' undergone a
major merger or significant mass jump is calculated as a function of
lookback time and interval.  The fraction of three- and four-body mergers is 
also studied.
\end{abstract}
\end{frontmatter}

\section{Introduction}

Due to their immense size, galaxy clusters can be easily identified 
with their dark matter halos, and their clustering and number counts
can be reliably predicted by dark matter simulations.
However in order to find or ``weigh'' galaxy clusters observationally,
assumptions of dynamical equilibrium and less understood astrophysics
usually come into play\footnote{Weak gravitational lensing
measures mass without equilibrium assumptions, but only in projection
(see e.g.~Refs. \cite{lens}).}.
Since halos form via the accretion and mergers of smaller units, and for
clusters this process is occurring to the present time, such assumptions
need to be specified precisely.  For example, how well a galaxy cluster is
described by equilibrium properties depends upon whether it has recently
undergone a major merger (where ``major'' means disrupting equilibrium)
and upon the specific methods used for its detection and mass measurement.
Thus even if one is only interested in galaxy clusters as ``test particles''
or peaks in the density field, the history of cluster growth is important in
order to make contact with observations.
Cluster growth histories are also important for understanding other cluster
properties.  For example, several observational phenomena (see below) are
associated with clusters which have recently undergone mergers.
A related question is which cluster properties have the least
sensitivity to cluster assembly histories.

In this paper we study the assembly history and degree of virialization
of high-mass halos in large N-body simulations.
A cluster assembly history can be characterized either by events with
specific occurrence times, such as mergers or large mass changes, or
by properties of its entire history, e.g.~a parameterization of
the mass as a function of time.
We calculate several such quantities for a statistically significant sample
of halos.
We consider several popular formation time definitions and cluster history 
parameterizations and statistics.
We also calculate the fraction of ``recently'' merged galaxy clusters as a
function of redshift, back to $z\sim 1$.

The definitions of both ``merger'' and ``recently'' depend on the cluster
property of interest and we explore several choices.
We also find the fraction of galaxy clusters which have had a recent
large mass gain (including accretion) for several choices of interval and
two final/initial mass ratios.  
Merger histories can be reliably extracted from N-body simulations, and
as such these recently merged fractions are implicit in earlier work.
However, it can be difficult to obtain specific numbers from the
literature for $\Lambda$CDM models, especially if one has a particular
relaxation time in mind.  In part this is because previous studies
of different quantities at different times have been published over
several years, often using different cosmological models.
Here we compute several of these quantities for a much larger sample
than used in earlier work, for $\Lambda$CDM cosmologies, and present them
in a homogeneous manner in the hope that this will be a useful reference for
the community.

The outline of the paper is as follows.
Section \ref{sec:background} is a review, including pointers to earlier
work on formation times and examples of observed merger phenomena.
Section \ref{sec:methods} describes the simulations and methods.
A reader interested primary in the results can skip directly to
Section \ref{sec:results}, which has comparisons and distributions of some
cluster formation properties, and the fractions of clusters
which have recently merged or had a large mass increase, as a function of
time.  Three and four-body major mergers are also studied.
Finally, Section \ref{sec:conclusions} presents our conclusions.

\section{Background} \label{sec:background}

The growth of structure by mergers and accretion is key to the
hierarchical paradigm of structure formation, and thus mergers
and mass gains have been studied intensively. 
Previous work on cluster formation histories includes
Refs.~\cite{TorBouWhi97,Tor98,Col99,GotKlyKra01,Zha,RowThoKay04,TKGK,Bus03}.
While cluster assembly is a complex and ongoing process in hierarchical
models, it is often useful to have some measure of when the cluster
``formed''.
Refs.~\cite{GotKlyKra01,RowThoKay04,TKGK}, each with 10-20 clusters,
considered formation time definitions including the redshift, $z_{\rm jump}$,
of the most recent large $\Delta M$ over a short time.
Both \cite{RowThoKay04,TKGK}
also found a characteristic formation time $z_f$ associated with 
the entire cluster growth curve using the parameterization of 
Ref.~\cite{Wec02}. This parameterization works extremely well for galaxy sized 
halos and correlates with other properties such as concentration.  For galaxy
clusters, Ref.~\cite{TKGK} introduced a generalization to help better
match the more recently active formation histories of galaxy clusters.
Ref.~\cite{Zha} had a sample similar to ours and found a
``turning point'' time where halos went from a quickly growing phase
to a more slowly growing phase, this turning point was correlated with 
concentration.  There are other formation times
considered in the literature.  For instance Ref.~\cite{vdB02}
found the average mass accretion history of halos generated by
extended Press-Schechter.
Ref.~\cite{RowThoKay04} also looked at the amount of mass gain coming from
large $\Delta M$ ``jumps'' as fraction of cluster mass.
We consider these properties and their distribution for our large
sample of over 500 clusters for a $\Lambda$CDM $\sigma_8 = 0.8$ model
in \S4.

Our second set of results is the frequency of recent mergers and 
recent large mass gains for some fixed lookback time and definition of
``recent.''  Analytic estimates 
of merger rates, in many cases combined with simulations, include
those by Refs.~\cite{analytic,Tor98,GotKlyKra01,CohBagWhi01} 
(see also the extension
reported in \cite{Nic02}).
The previous work closest to the merger counts considered here is by
Ref.~\cite{GotKlyKra01} who found the
major merger distribution as a function of redshift for a system which had
11 clusters at $z=0$ (with a merger defined as an increase in halo
mass such that $M_f/M_i > 1.33$) 
and that of Ref.~\cite{CohBagWhi01} which also studied
the recently merged population, but for a smaller number of clusters
and not in as much detail.  It should be noted that merger rates and
the number of clusters which have recently undergone a merger are
slightly different, it is the latter we study here.

Mergers can alter X-ray temperatures, luminosities, cluster galaxy light, 
cluster galaxy velocities and Sunyaev Zel'dovich
(SZ; Ref.~\cite{SZ}) ``flux''\footnote{Several simulations with different
heating and feedback prescriptions have found small merger induced scatter
in the integrated SZ flux relation
(Refs.~\cite{WhiHerSpr02,Mot05,Koc05} although there are
examples where a $\sim 50\%$ increase in total SZ flux for one sound crossing
has been seen \cite{Sar05}).
Significant scatter in the mass-flux relation instead appears to be
dominated by that due to line of sight projection \cite{WhiHerSpr02}.}.
Observational consequences of merging clusters of galaxies have been
studied analytically and with numerical simulations,
e.g.~Refs.~\cite{numer,RicSar01,RowThoKay04}.  In these, 
the time scale for major disturbances of clusters due to 
mergers seems only weakly dependent upon the impact parameter (the 
magnitude of the disturbance has a stronger dependence, e.g.~see
Refs \cite{RicSar01}).  We do not consider impact
parameters below.  The errors
induced by ignoring merger effects (on  X-ray luminosities and 
temperatures) for cosmological parameter estimates 
has been considered in \cite{RanSarRic02}\footnote{They used an analytic model
\cite{SomKol99} to get a merger history,
in principle our calculations here could also be used for this purpose.}.
More generally, unaccounted for mergers can disguise cluster masses
and alter survey selection functions.

Observationally, some signatures of mergers can be used to flag unrelaxed
clusters.  For instance substructure, the presence of multiple peaks in
the cluster surface density on scales larger than the constituent galaxies,
has been studied extensively.
Several different observational methods have been used to quantify the
amount of substructure in X-ray clusters
(see the review \cite{Buo02} and references therein).
The fraction of clusters exhibiting substructure has been found to be
significant in several surveys.
For example, Ref.~\cite{JonFor99} visually found substructure
in 41\% of 208 Einstein IPC images, Ref.~\cite{Moh95} measured the
emission weighted centroid variation for 65 Einstein clusters to get a
substructure fraction of 61\% (see also \cite{moresub} for related work)
and Ref.~\cite{Sch01} found, using three different indicators,
substructure in $52\pm 7\%$ of 470 clusters in the ROSAT all-sky survey.
Substructure has also been searched for in other wave-bands.
For example, substructure was detected in over 80\% of 25 low richness
2dFGRS clusters (\cite{Bur04}; see their references for other
studies).
Following Ref.~\cite{TsaBuo96}, substructure has been considered as
a means of constraining cosmology, e.g. in \cite{Jel05} using
40 Chandra clusters over a range of redshifts
(see also \cite{struccos} for relevant simulations).
A meaningful comparison of our calculated number of ``recently merged''
clusters to the number of observed galaxy clusters with substructure
depends on relaxation times.  These times depend upon the substructure
measurement in question and are still not well estimated.
For instance, relatively small sub-clumps may take a different amount of
time to disappear depending on their density, see \cite{Buo02} for
further discussion.  

Radio halos and radio relics have been proposed as another merger indicator
(see the reviews \cite{Fer04} and \cite{Sar04b}).
Clusters hosting radio halos and relics tend to show other signs of 
merger activity, however
not all clusters with other indicators of recent mergers show these
radio signals.  In a complete sample \cite{GioFer02} at the NRAO VLA
Sky Survey surface brightness limit, only 5\% of clusters had a radio halo
source and 6\% had a peripheral relic source, however in clusters
with X-ray luminosity larger than $10^{45}$ erg s${}^{-1}$, 
35\% had radio halos or relics.  In the ROSAT sample considered
above by Ref.~\cite{Sch01}, 53 clusters of the 470
had radio halos or relics, most of these were high luminosity
($L > 4.0 \times 10^{44}$ erg s${}^{-1}$).  It has been
proposed that more radio halos and relics 
will be found in lower mass clusters as sensitivities improve
(e.g.~ Ref.~\cite{Sar05}).  The relaxation
time for radio halos and relics is not known.

The absence of mergers (i.e. a relaxed cluster) has been suggested as
a requirement for cool core clusters (e.g.~Refs.~\cite{cool}).
The fraction of cool core clusters
has been measured for various samples, e.g.~Ref.\cite{Bau05}
found with Chandra that 55\% of 38 clusters in
the ROSAT Brightest Cluster Sample (with $z \sim 0.15-0.4$)
show mild cooling (associated with a cooling time less that $10\,$Gyr) 
and 34 \% show signs of strong cooling (a cooling time less than $2\,$Gyr).
There are some simulations that suggest however that mergers may help 
form cool galaxy cluster cores (e.g.~see \cite{Mot04}),
also 5/22 of the ROSAT cooling flow clusters have significant substructure
\cite{Sch01}.

There are many more phenomena associated with
mergers which are even more complex
(see Sarazin's list of questions about merging clusters of
galaxies \cite{Sar04}).  The examples of merger indicators
reported above have had their occurrence rates measured in some of 
the largest cluster samples available.

There are some differences between mergers and any form of mass
gain\footnote{A modified version of the PS formalism that differentiates
between (instantaneous) major mergers and accretion has been developed
in Refs.~\protect\cite{mer-ac}.}.
Clusters which have gained mass primarily via accretion might be expected
to be more relaxed than those who reach the same mass via a major merger.
For smaller halos those with a relatively large mass gain over a short
time are more clustered than halos of the same mass but without this mass
gain \cite{ScaTha03}.
The subset of recently merged halos, in contrast, does not appear to be
so biased \cite{Per03}.  The authors of Ref.~\cite{ScaTha03} conjecture
that objects in denser regions tend to have more nearby material to
accumulate, resulting in a bias, while the same is not true
of recent mergers.

\section{Methods} \label{sec:methods}

To investigate these questions we used two N-body simulations run with a
{\sl TreePM\/} code \cite{Whi02}.  Each simulation evolved $512^3$ particles
in a periodic cubical box, $300\,h^{-1}$Mpc on a side, for a particle
mass of $1.67 \times 10^{10}\,h^{-1}M_\odot$.
The models had $h=0.7$, $\Omega_m=0.3$, $\Omega_\Lambda=0.7$ and
$\sigma_8=0.8$, 1.0.
The two normalizations span the observationally preferred range.
Outputs were dumped at equal intervals\footnote{This is the same time
interval considered by Ref.~\cite{RowThoKay04} and shorter than that
considered by Refs.~\cite{GotKlyKra01,TKGK}, though the latter groups
still had their time spacings well within the lower limits for a
merging time scale.} of conformal time, $\delta\tau=100\,h^{-1}$Mpc
(comoving), starting at $z=2$.
Other parameters and details of the simulations can be found in
\cite{YanWhi04}.

For each output groups are identified with a friends of friends algorithm
(FoF; \cite{Dav85}) using a linking length $b = 0.15$ in units
of the mean inter-particle spacing.  These groups correspond roughly to
all particles above a density of $3/(2\pi b^3)\simeq 140$ times the
background density.
We did not consider other group finders, a comparison between HOP
\cite{EisHut98} and FoF merger fractions for a smaller sample was
considered in \cite{CohBagWhi01}.
In that case there were both fewer recently merged clusters and fewer
clusters altogether for the FoF case.
See also Ref.~\cite{ScaTha03} for a comparison of HOP and FoF mergers
and accretion for smaller mass halos.
They estimate what minimum mass fraction in accretion is robust for
different group fractions and find that a 20\% mass increase (for a shorter
length of time and smaller halos than we consider) is outside the ``noise''
region for the FoF vs.~HOP group finders.

For each group and output time we compute a number of properties, including
the mass, velocity dispersion and potential energy of the group.  All
properties are computed using only group members.
Masses henceforth refer to friends of friends masses $M_{\rm fof}$ unless
otherwise stated.  To get $M_{200}$ or $M_{\rm vir}$ conversion is needed
\cite{Whi01} -- for a fitting formula see \cite{HuKra03}.
{}From the simulation we find $M=1.2\;M_{200}$ at $z=0$, and
$M=0.98\;M_{200}$ at $z\sim 1$.
We consider halos to be clusters if $M\geq 10^{14}\,h^{-1}M_\odot$ 
(i.e.~5971 or more particles); at $z=0$ there are 909 clusters in the
$\sigma_8=1.0$ box and 574 in the $\sigma_8 = 0.8$ box.

For the formation time studies, most of the definitions are straightforwardly
derived from the cluster histories.  We track the properties of all of the
progenitor halos with more than 64 particles for all of our clusters.  From
these histories it is straightforward to find the mass of the largest
progenitor as a function of time.
To find mergers, the two largest predecessors are found for each cluster at
an earlier time.  The final cluster is said to have had a (major) merger
between the earlier time and the later time if the ratio of particle numbers
going from both of the predecessors into the final cluster is above some
minimum\footnote{This was in part to try to minimize the effect of particles
which were not actually bound being taken to be part of a cluster's 
predecessor.}.  In the case where the time interval of interest does not
coincide with a dump time for the data, the two closest times
bracketing the time of interest are taken, and the merger fraction is
linearly interpolated between these (starting from the time closest to the
time of interest).  One relaxation time we consider is the crossing
time, in this case we start at the earlier time and interpolate between
final times.

We checked our definitions of mergers in two ways. Instead of
using the mass of only progenitor particles which actually go into the
final halo in the mass ratio, one can use the full mass of the
progenitor halos, this changes the merged fraction
by less than 5\% for the longer time intervals
and almost always less than 10\% for the short time intervals.
We prefer our definition as it might help counteract the inclusion
of many unbound particles in halos due to our FoF group finder.
Secondly, we identify mergers
by looking at the initial and final halos and comparing mass ratios
of the initial halos, without tracking their intermediate behavior until they
merged to the final halo.  A concern was
that two predecessors with an initial mass ratio of e.g. 1:5 might first
become a mass ratio of 1:10, say, if the largest progenitor grew significantly
in mass relative to the second largest, before merging.  Looking at
intermediate times, the mass ratio of predecessors is
differs for the two definitions, but mostly for almost equal mass
predecessors.  Once the predecessor mass
ratio drops to $\sim$ 1:3 the scatter between the two ways of 
defining predecessor mass ratio gets quite small.  Using the 
more accurate definition based on intermediate time steps produces
a slightly larger number of recently merged clusters, again less than 5\%.
That is, our approximation in some cases underestimates the number of
mergers by a small amount.

We vary the mass ratios and time intervals, and consider intervals
equally spaced in real time, in light crossings (scaling as $a$) and
in terms of ``crossing times''
(scaling as $(G\bar{\rho})^{-1/2}\sim a^{3/2}$).
We call the lookback time the time at which the cluster is observed, and the
relaxation time the interval prior to the lookback time within which the
merger or mass jump (``large $\Delta M$'') has or hasn't occurred.
Intervals of 100, 300 and $600a\,h^{-1}$Mpc light crossings correspond
to approximately 1, 3 and $6\times a/2\,$Gyr.
We call these $\delta\tau = 100, 300, 600\,h^{-1}$Mpc in the following.
Although the simulations go back to $z\sim 2$, we only plot back to
lookback times where the statistics have any constraining power.

We also consider three- and four-body merger fractions, since such
systems have been observed (see, e.g.~Refs.~\cite{threebody}).
We define three-body mergers as the case
when the second and third largest predecessors each contribute
to the final halo at least 20\% of the mass contributed
by the largest predecessor.
Similarly, four-body mergers require the fourth largest predecessor to also
contribute at least 20\% of the mass contributed by the largest predecessor.

\section{Results} \label{sec:results}

\subsection{Formation times and properties}

\begin{figure*}[htb]
\begin{center}
\resizebox{5.5in}{!}{\includegraphics{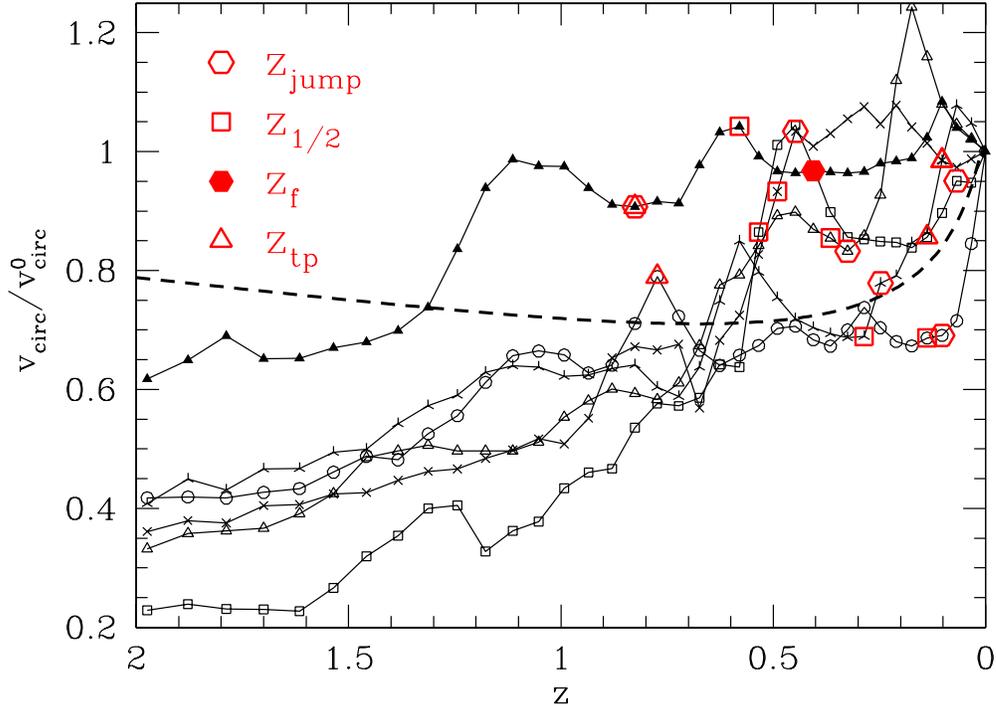}}
\end{center}
\caption{The evolution of the peak circular velocity of the most massive
progenitor for 6 massive clusters (symbols connected by solid lines)
compared to spherical top-hat collapse of a cluster virializing at $z=0$
(dashed).  The cosmological model is our $\Lambda$CDM $\sigma_8 = 0.8$
model.  The large symbols indicate various formation times discussed in
the text, as indicated by the legend.
Only the cluster denoted by the filled circle has a positive $z_f$.}
\label{fig:sthc}
\end{figure*}

A classic paradigm for cluster assembly histories is the spherical top-hat 
collapse model.  In this model a uniform, overdense region ``breaks away'' 
from the expansion of the universe, evolving as a self-contained positive 
curvature universe.  This toy model is often used to motivate the threshold 
density for virialization or to estimate formation times
\cite{LiddleText}.
The evolution of such perturbations does not closely resemble the rich
structure seen in the formation of massive halos in N-body simulations,
as noted by many authors referenced above, however it is interesting to
see how it performs nonetheless.  
We show in Fig.~\ref{fig:sthc} the evolution
of the peak circular velocity ($M\propto v_c^3$) of the most massive 
progenitor for 6 clusters, chosen at random from amongst the most massive 
($M > 6 \times 10^{14} h^{-1} M_\odot$) in our
$\sigma_8=0.8$ simulation.  Since the mass in the spherical top-hat model
is difficult to compare to simulations (it is constant and has uniform
overdensity), we use peak circular velocity.
The peak of the circular velocity curve, $v_c^2\equiv GM(<r)/r$, is computed
for each progenitor using the minimum of the cluster potential as a center
for defining $r$.
The dashed line shows the evolution predicted using the spherical top-hat
collapse model for a cluster which virializes at the present.  We see that 
the spherical infall description has a cluster's peak circular
velocity growing most quickly right 
before virialization, while clusters forming in a cosmological
simulation have more of a steady growth over time\footnote{For the spherical 
top-hat collapse model, the cluster mass is constant and the density/radius
change with time.  In physical coordinates the perturbation first expands
(falling $v_c$) and then collapses (rising $v_c$).}.
We have also marked on the plot four of the formation times defined and
discussed below, $z_{\rm jump}$, $z_{1/2}$, $z_f$ and $z_{\rm tp}$.

\begin{figure*}[htb]
\begin{center}
\resizebox{5.5in}{!}{\includegraphics{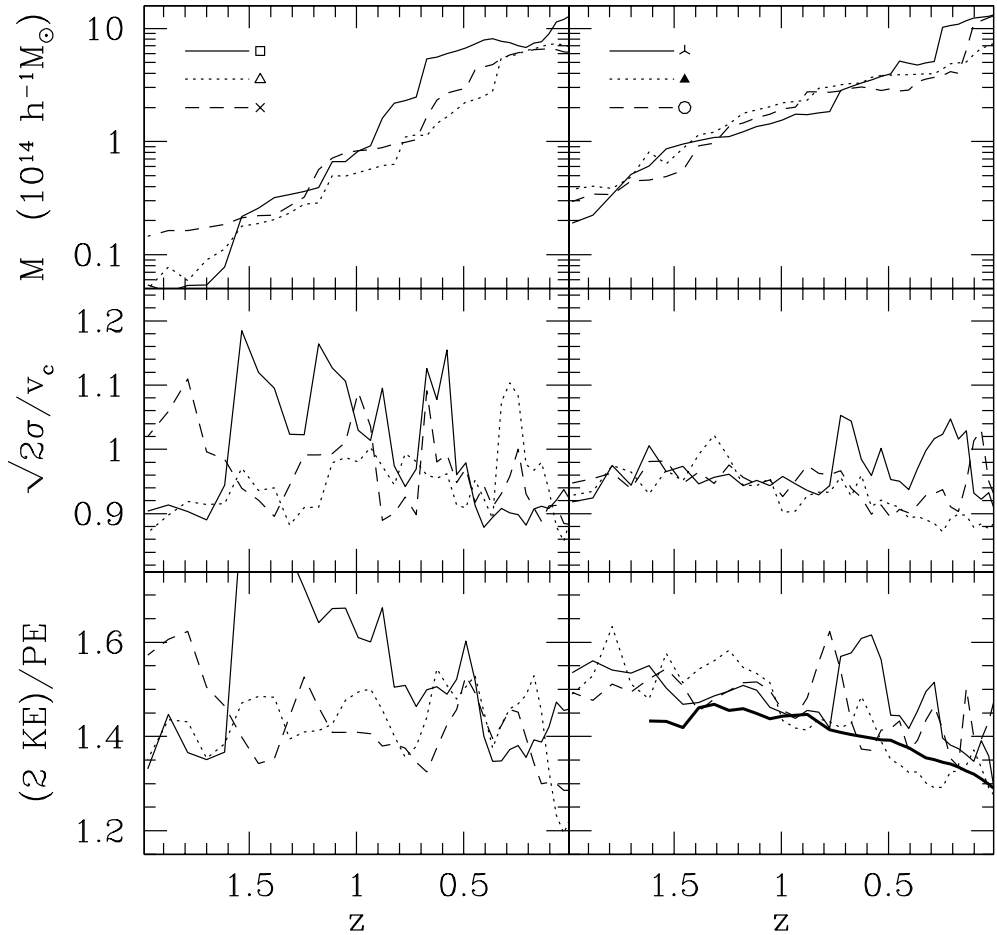}}
\end{center}
\caption{A view of cluster formation for the
same 6 massive clusters as in Fig.~\ref{fig:sthc}
(see inset legend for correspondence).
The left and right columns show 3 clusters each.
The top panel shows the mass accretion history since $z=2$, the
middle panel shows the ratio of velocity dispersion to circular velocity
(which has been used previously as a proxy for degree of virialization)
while the lower panels show the virial ratio, $2{\rm KE}/{\rm PE}$, as a
function of time (see text).
In the lower right panel, the heavy solid line is the virial ratio averaged
over all clusters with $M>10^{14}\,h^{-1} M_\odot$ at the given
redshift.  The first point, $z\simeq 1.6$, has 3 clusters above the
threshold.}
\label{fig:virial}
\end{figure*}

A different view of the formation of clusters is given in
Fig.~\ref{fig:virial} which focuses on the same 6 clusters as
Fig.~\ref{fig:sthc}.
In the top panels we show the mass of the largest progenitor as a function
of time. In the middle panel we show the ratio of the 1D velocity dispersion
to the peak circular velocity -- this has been used as a proxy for the degree
of virialization of the clusters \cite{KneMul99}.
In the lower panel we show the relation between the total kinetic and
potential energy explicitly.
Both energies have been computed using only the particles in the FoF group, 
but no proxy has been used, unlike the middle panel.

Even focusing only on the most massive progenitor, rather than the full
merger tree, the upper panels show the rich phenomenology of cluster
formation.
There are several obvious mass jumps interspersed between periods of
relatively smooth accretion. The lower panels show that the cluster
is disturbed by mergers and large mass jumps with strong departures
from the expected virial relation $2{\rm KE}\simeq {\rm PE}$.
The departures last for several of our output times.
The ``standard'' vacuum virial relation $2{\rm KE}\equiv {\rm PE}$ is
not satisfied by these clusters even in their quiescent phase.
The ratio $2{\rm KE}/{\rm PE}$ is typically larger than unity due to the
steady accretion of material onto the cluster\footnote{This can be
thought of as a surface pressure term which modifies the virial
relation, boosting the kinetic energy.}\cite{ColLac96,KneMul99} 
and there is an overall decline
in $2{\rm KE}/{\rm PE}$ with time. 
Averaging over the entire cluster sample we find for objects of fixed
mass the KE/PE ratio decreases with decreasing redshift.
For objects at fixed redshift the ratio increases with increasing mass.
For objects at fixed number density the ratio decreases with decreasing
redshift.
For fixed mass cut or number density the decline is steeper
below $z=1$ than above $z=1$ and within errors the decline below
$z=1$ has the same shape for all samples. 
We hypothesize that this decline in KE/PE as due to the cessation of
structure formation due to $\Lambda$ domination below $z\sim 1$
(plus a small shift in the mass of objects currently undergoing rapid
accretion).  We will see a similar drop in merger activity below 
$z\sim 1$ in the next section.

We attempted to further quantify the relaxation of clusters to the
background virial relation as a function of time since last major
disturbance (c.f.~Ref.~\cite{RowThoKay04}), but we found the scatter was
too large for us to draw robust conclusions.

Even though cluster formation is a complex process it is sometimes useful
to attempt to describe it with a single ``formation time''.  Any such
compression of information must be imperfect, and to some extent arbitrary.
Depending on the phenomenon of interest different times may be more or
less appropriate.  For this reason several definitions in the literature
exist for the ``formation time'' of a cluster.  We calculate several of
these and their distributions for the $\sigma_8=0.8$ sample of 574 clusters
below.

\begin{figure*}[htb]
\begin{center}
\resizebox{5.5in}{!}{\includegraphics{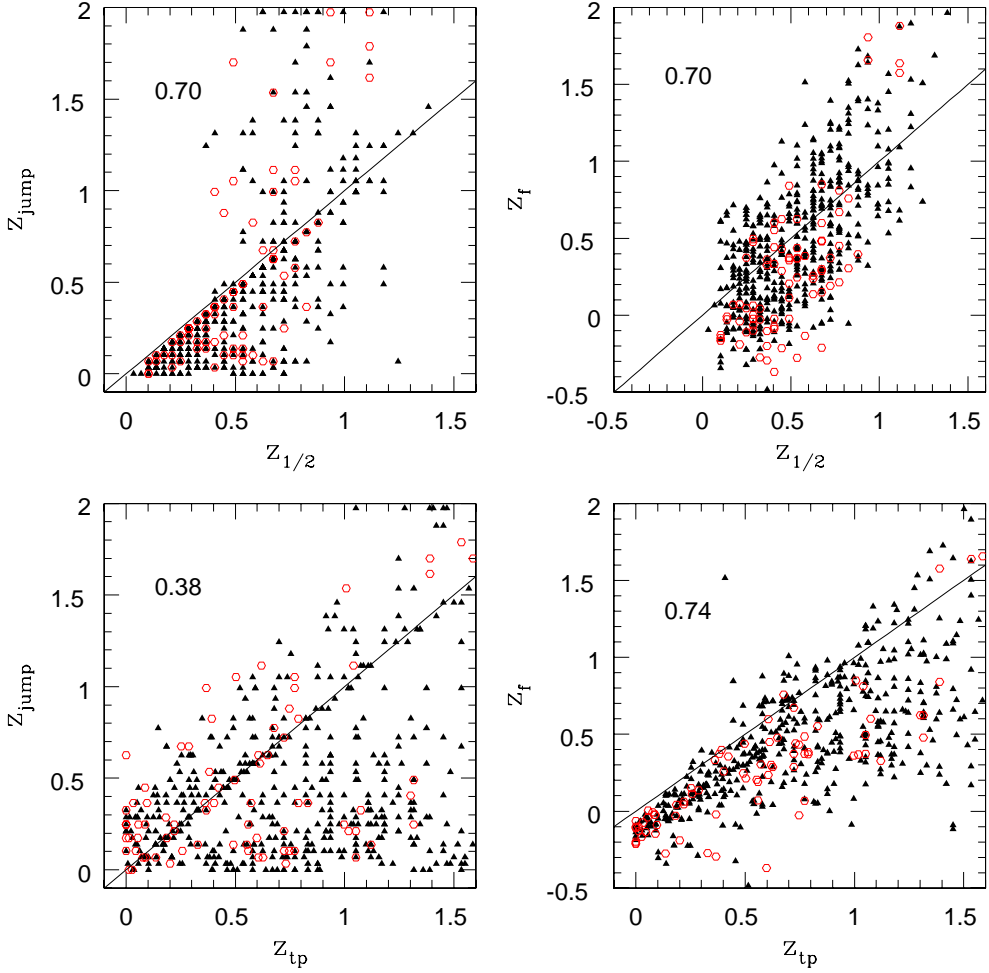}}
\end{center}
\caption{Top left: $z_{1/2}$ vs.~$z_{\rm jump}$.  The 13 clusters with no
large $\Delta M$ in the simulations are not shown as they have
$z_{\rm jump}>2$.  Top right: $z_{1/2}$ vs.~$z_{f}$.
Lower left: $z_{\rm tp}$ vs.~$z_{\rm jump}$.  Lower right: $z_{\rm tp}$ 
vs.~$z_f$.  The red circles correspond to clusters with
$M>3\times 10^{14}\,h^{-1}M_\odot$.  The correlation coefficient, $r$, for
each case is shown at upper left.
}
\label{fig:formtime}
\end{figure*}

The first definition we consider is when a cluster had its most recent large
$\Delta M$, i.e.~$M_f/M_i\geq  1.2$ in an interval $\delta\tau=100\,h^{-1}$Mpc.
In this section a mass ``jump'' refers to a mass gain of at least this
much.  For any cluster $z_{\rm jump}$ is the redshift of the most recent 
such jump.  In our sample 13 clusters had no large mass jumps at any time
after $z\sim 2$, two of these clusters are more massive than
$3\times 10^{14}\,h^{-1}M_\odot$.  Ref.~\cite{RowThoKay04} 
found that 2 out of their sample of 20 did not have any large mass jumps.
Another common definition is when the cluster has reached at least half
its mass, for this time there is an analytic formula
\cite{NusShe99} which was tested in (and improved using)
simulations in Ref. \cite{SheTor04}.
We define $z_{1/2}$ as the earliest output time when the cluster had at least
half of its mass.  The relation between these two definitions is
shown in the upper left hand panel of Fig.~\ref{fig:formtime}.
There is a large amount of scatter between the two definitions but
there is a striking transition at $z = 0.5$.  For clusters which have
not reached half of their present day mass before $z = 0.5$, almost
all have had a large $\Delta M$ jump after reaching half-mass, i.e.~they have
at least half of their mass before they have their last large mass jump.
However, for clusters which attained $z_{1/2}$ before redshift $z = 0.5$
(286 clusters),
slightly under one third (80) had their most recent mass jump
even earlier (i.e.~$z_{\rm jump} > z_{1/2}$).
An intuitive reason for this is that clusters which reached half of their
mass early on have not been gaining mass quickly and so are less
likely to have large mass jumps after $z_{1/2}$.
This general shape persists if one considers some other ``formation time''
such as $z_{1/3}$ or $z_{3/4}$ (with obvious definitions), the $z$ value
where $z_{\rm jump}$ starts becoming larger (that is, earlier) than 
$z_{1/3} (z_{3/4})$ is larger (smaller) than that for $z_{1/2}$.
A larger fraction (12/33) of the clusters with mass greater than 
$3\times 10^{14}\,h^{-1}M_\odot$ and $z_{1/2}>0.5$ had $z_{\rm jump}>z_{1/2}$.
If one is interested in formation times in order to find undisturbed 
clusters, for lower redshifts it appears that $z_{\rm jump}$ may be a more
conservative estimate.

A third definition of a cluster formation time is when the potential well of 
the object becomes deep enough to be considered ``a cluster''.   We take this,
somewhat arbitrarily, to be when the object has reached 
$10^{14}\,h^{-1}M_\odot$, and call this $z_{14}$.

All of these definitions rely on special events in the mass accretion
history.  An alternative is to use the whole history of the most
massive progenitor.  While this is less information than in the entire tree,
it provides a global view of the formation process.  We consider three
fits used in the past for simulations.
For galaxy sized halos, Ref.~\cite{Wec02} found that the mass accretion
histories could be fit by
$M(a)=M_0 e^{-2a_f z} = M_0e^{-2z/(1+z_f)}$, and that $z_f$ correlated
well with other cluster properties, i.e.~concentration.
Ref.~\cite{TKGK} found that for clusters they could obtain better fits
if they generalized the function to
$\widetilde{M}(a)=M_0 a^p e^{-2\widetilde{a}_f z} = a^p M(a)$, where 
$\widetilde{a}_f = 1/(1+\tilde{z}_f)$.  A third form was proposed by
\cite{Zha}, who fit the the mass accretion history to the form
$M_i/M_{\rm tp}=f\left(\rho_{\rm vir}(z_{\rm tp})/\rho_{\rm vir}(z_i)\right)$
where $\rho_{\rm vir}(z)$ is the virial density at redshift $z$ computed from
spherical top-hat collapse and $f$ interpolates between two power-laws
\cite{Zha}.
The latter authors found that halos generally transition from
a period of rapid accretion to a phase of slow accretion and used the
transition redshift, $z_{\rm tp}$ as a proxy for ``formation'' time.

We fit these three parameterizations to our sample,
estimating $z_f, \tilde{z}_f,z_{tp}$ through a least squares
fit of $\ln(M_i/M_0)$ against $z_i$, with all points being equally 
weighted.  The values of $z_f,\tilde{z}_f$ are allowed free range,
however $z_{\rm tp}$, due to its definition, is constrained to the
range $0<z<2$ where we have data\footnote{Also, we fit $M$ not 
$M_{\rm vir}$ but this should have little effect since
$M_{\rm vir}\simeq 1.08\,M$ at $z=0$ and $ \simeq 1.10\,M$ at 
high $z$.}.
We compare this formation time $z_f$ with $z_{1/2}$ at upper right in 
Fig.~\ref{fig:formtime}; they are correlated (as also found by 
Ref.~\cite{TKGK}).
We show in Fig.~\ref{fig:formtime}, bottom, the comparison of this ``turning
point'' time $z_{\rm tp}$ with $z_{\rm jump}$ and $z_f$ above.
The largest correlation shown is between $z_{\rm tp}$ and $z_f$, as shown in
Fig.~\ref{fig:formtime}.

\begin{figure*}[htb]
\begin{center}
\resizebox{5.5in}{!}{\includegraphics{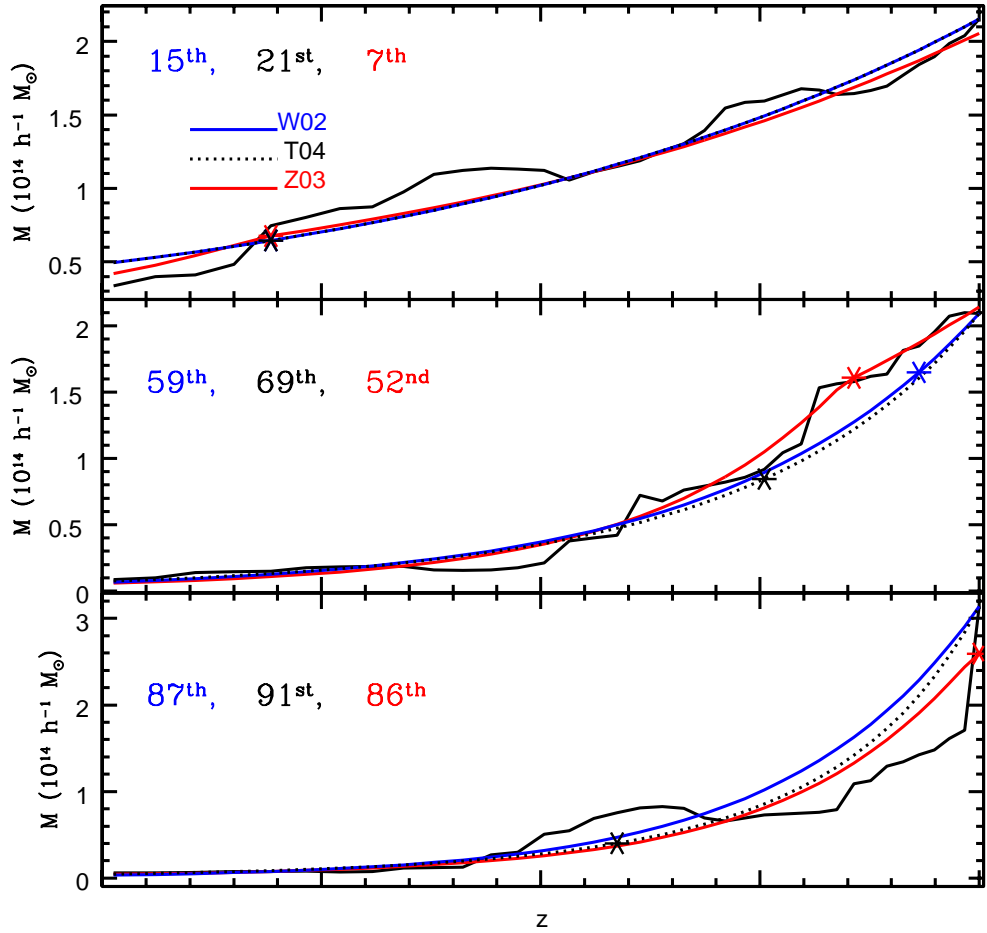}}
\end{center}
\caption{Examples of parameterized fits to three clusters, one of the
better fits (top), less good (middle) and worst (bottom).  The solid black
line in each panel is the history of the most massive progenitor, the
solid blue/dotted/red lines are the fits of Refs.~\cite{Wec02}, \cite{TKGK}
and \cite{Zha} respectively.  For the top cluster the fits of 
Refs.~\cite{Wec02} and \cite{TKGK} coincide.
The stars are the associated formation times (for the bottom cluster the 
formation time of \cite{Wec02} is in the future).  We list the percentiles
in goodness of fit for each of the fits at upper left.}
\label{fig:threechi}
\end{figure*}

Although these parameterizations do describe general trends in the cluster 
mass histories, good fits were not found for all the cluster 
histories.  (The best fits using our least squares criterion was for 
the models of Ref.~\cite{TKGK}, which have an extra parameter.  However,
similarly to \cite{TKGK}, we found that many of the clusters had either 
$p = 0$ (278/574) or $\tilde{a}_f = 0$ (119/574).)
This made results sensitive to the $z$-range used and the 
interpretation of the fit coefficient difficult.  To illustrate the
difficultly, we show in Fig.~\ref{fig:threechi} examples of fits for 3
clusters.

In more detail, for the fit for $z_f$ (introduced by Ref.~\cite{Wec02}),
we investigated the dependence on the range of $z$ used in the fit.
Ref.~\cite{TKGK} fit their $z_f$ to histories for $0\le z<0.5$ and 
for $0\le z<10$ and found that the results were consistent.  We found 
extremely large variations comparing  $0\le z<0.5$ and 
$0 \le z <2.07$.  Our fits may have differed in success from
theirs for two reasons.  Our clusters tend to be more massive (theirs
were between $5.8 \times 10^{13} h^{-1} M_\odot \leq M_{180 \rho_b} \leq
2.5 \times 10^{14} h^{-1} M_\odot$, which are even lower masses in terms
of our mass definition), as they note, less massive halos tend to have
better fits in general.  Secondly, our sample is a lot larger, allowing us
to sample a broader range of behavior.
Even some clusters which gained more than 90\% of their mass in the
range we consider have $z_f$ which depends sensitively on the fit range.
This is a simple consequence of the fact, which can be seen above,
that the functional form does not
well describe the shapes of the mass accretion histories of these objects.
Finally we note that for clusters with recent mass jumps the fit prefers
$z_f<0$.  As it is very plausible to say that these systems are still
in the process of forming we report these values above without renormalization.

The other two parameterizations also had difficulties with some subset
of the clusters.  It should be noted however that our redshift range was
smaller than \cite{TKGK} and that they did not report robustness of
the $p \ne 0$ fits to smaller $z$ ranges.  Ref.~\cite{Zha} actually
considers a sample of similar size to ours for their fit, and large
scatter is evident around the parameterized fit in their results as well,
they find a useful correlation between $z_{tp}$ and concentration nonetheless.

\begin{figure*}[htb]
\begin{center}
\resizebox{5.5in}{!}{\includegraphics{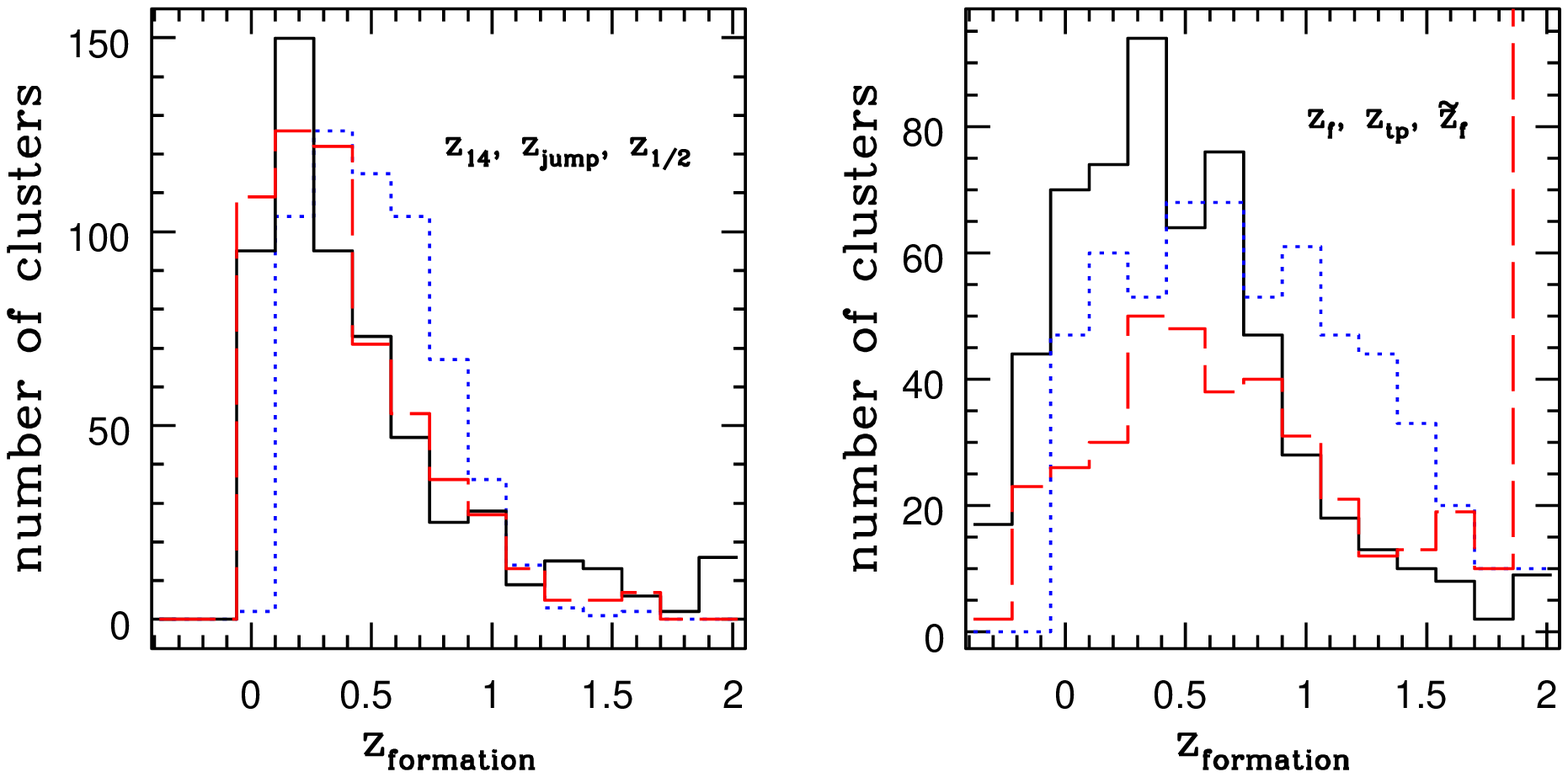}}
\end{center}
\caption{Distributions of different formation time definitions 
for all clusters in the $\sigma_8 = 0.8$ sample.
On left are the distributions for
$z_{\rm jump}$ (solid line), $z_{1/2}$ (dotted line) and
$z_{14}$ (dashed line).  At right, is the distribution for
$z_f$ (solid line), $z_{\rm tp}$ (dotted line) and $\tilde{z}_f$ (dashed
line). 
Note that some clusters have $z_f<0$.  See text for further definitions
of these formation times.}
\label{fig:formdist}
\end{figure*}

In addition to comparing the different formation times to each other,
we also use the cluster sample to get a snapshot of the
formation time distribution.
In Fig.~\ref{fig:formdist} we show the distributions of these
different formation times for all the clusters at $z=0$.
Clusters which had no large mass jump since $z=2$ are put in the
$z_{\rm jump}=2$ bin for completeness.

There is quite a spread in these formation times, however some trends are
clear.  By $z=0.2$ over 85\% of the clusters have had their last large
mass jump and over 81\% of them have formed according to the definition of
Ref.~\cite{Wec02}, while 88\% of them have at least half of their mass.  
The different definitions have different biases though, e.g., of the
redshift formation times shown, only the
definition of Ref.~\cite{Wec02} can go negative.

\begin{figure*}[htb]
\begin{center}
\resizebox{5.5in}{!}{\includegraphics{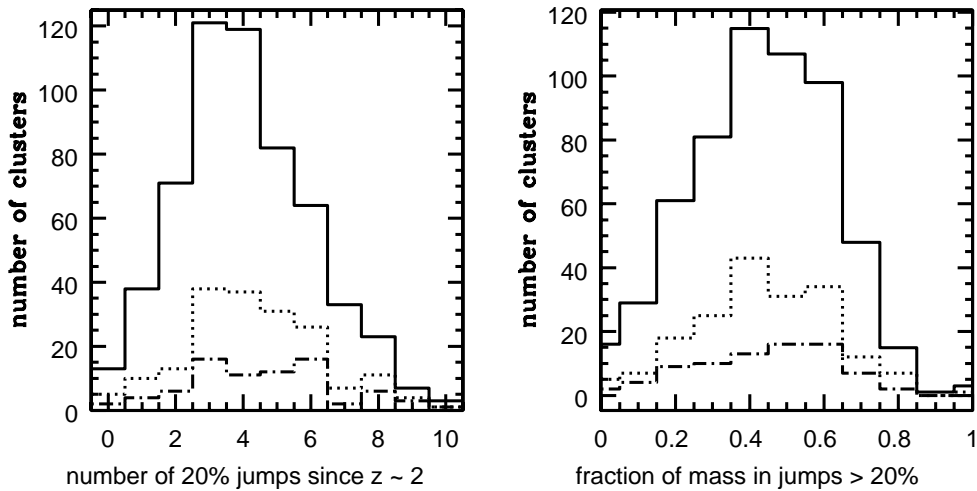}}
\end{center}
\caption{Left: Number of clusters with $N$ mass jumps $M_f/M_i\geq 1.2$
since $z=2$.  The three lines are for clusters with with $M>1$, 2, and
$3\times 10^{14}\,h^{-1}M_\odot$.  The average number (median number)
of jumps are 4.0(4),4.3(4),4.6(5) respectively.
Right: number of clusters with a given fraction of the total cluster mass 
obtained in mass jumps $M_f/M_i\geq 1.2$.}
\label{fig:jumps}
\end{figure*}

A major part of our difficulty in fitting smooth curves to the mass accretion
histories is the presence of significant mass jumps.  These jumps are
evident in the examples and are a general feature in cluster formation.
The statistics of these jumps are of interest.
Fig.~\ref{fig:jumps} gives the distribution of the number of times since
$z=2$ that the mass of each cluster changes by 20\%.
This is the same quantity considered by Ref.~\cite{RowThoKay04} for their
sample of 20 clusters more massive than $1.2\times 10^{14}\,h^{-1}M_\odot$
-- our sample is more than an order of magnitude larger allowing us to
more completely characterize the distribution.  We find that there is little
dependence on the cluster mass -- lines for clusters above 1, 2 and
$3\times 10^{14}\,h^{-1}M_\odot$ are shown.  The average (median) number of 
jumps for each of these samples is 4.0 (4), 4.3 (4), 4.6 (5), respectively,
with a wide spread.
The right panel of Fig.~\ref{fig:jumps} shows how much of the cluster
mass is gained in these events -- slightly less than half of the clusters 
(272/574) get at least half of their mass in jumps of 20\% of larger.
For the $M>3\times 10^{14} h^{-1}M_\odot$ subsample the fraction is larger:
41/79.

With a several large number of jumps common, it appears difficult
to find good smooth and simple parameterized fits for all of the histories.
However, given the correlations between the different fit ``formation times''
and between these and e.~g.~concentration, these fits may be very useful
approximations in a large number of cases. 

\subsection{Time dependence of recently merged or
large $\Delta M$ cluster fraction}

The above shows that the present day cluster population has a very wide
spread of formation times and frequent major disruptions.
For observational purposes, another question is: how many
clusters at a given redshift (``lookback time'') have had a large 
disruption recently (within a given ``relaxation time'').

\begin{figure*}[htb]
\begin{center}
\resizebox{5.5in}{!}{\includegraphics{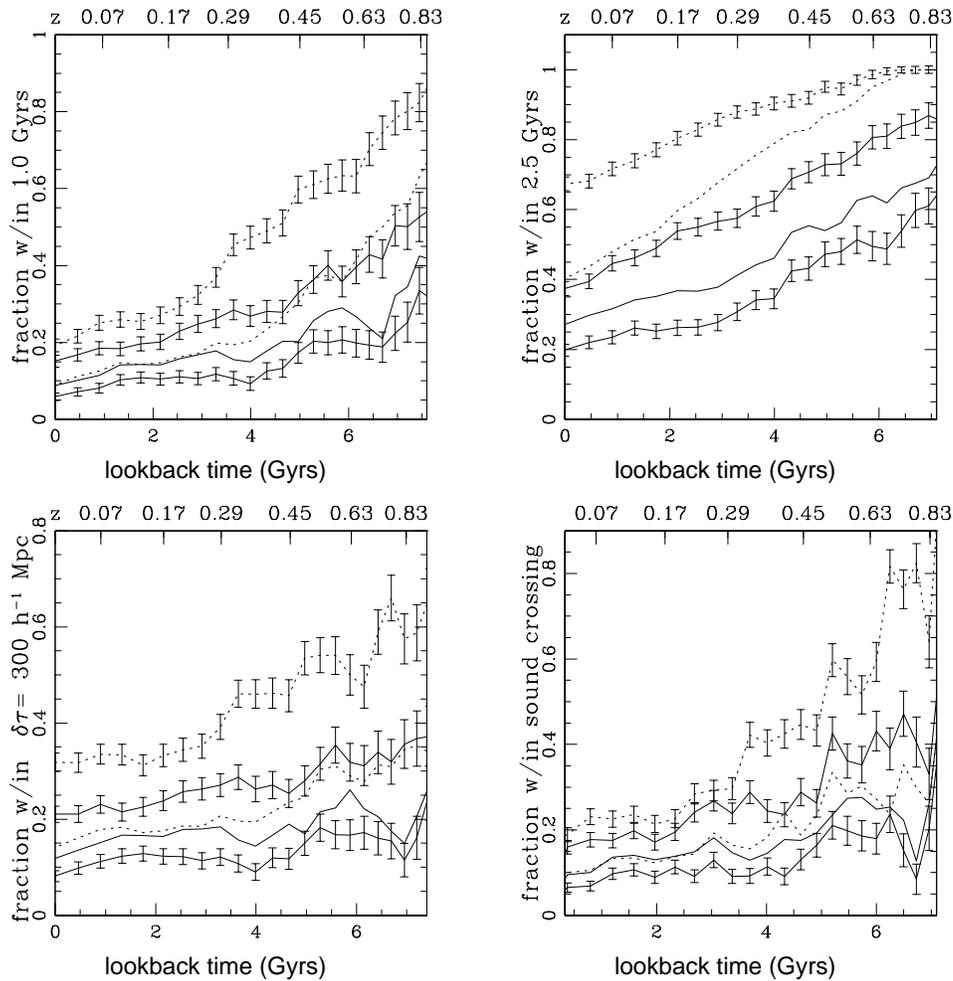}}
\end{center}
\caption{Top: Merger fractions in the previous $1.0\,$Gyr, $2.5\,$Gyr,
$\delta\tau = 300\,h^{-1}$Mpc and $t_{\rm cross}$, as a function of
lookback time to clusters, for $\sigma_8= 0.8$.  The solid lines give the
fraction of clusters ($M\geq 10^{14}\,h^{-1}M_\odot$) whose two largest
predecessors contribute numbers of particles with ratios of 10:1 or 
smaller, 5:1 or smaller and 3:1 or smaller (top to bottom) at given lookback 
times.  The dotted lines are the fraction of clusters which have
$M_f/M_i \geq1.2$ and 1.33.
For clarity (binomial) error bars are shown for only some of the lines.
The others are similar.}
\label{fig:merger1}
\end{figure*}

The fraction of clusters with a large mass change $\Delta M$
in a given interval is shown in Fig.~\ref{fig:merger1}, top, as the dotted 
lines. This is the total mass change, via mergers {\it or\/} accretion, within
the 1.0 and $2.5\,$Gyr time intervals.  The two lines show the fraction of
clusters with final/initial mass $ \geq 1.2$ and $1.33 $ (these ratios were
chosen by Refs.~\cite{RowThoKay04} and \cite{GotKlyKra01}), with the lower 
line for the higher mass ratio.

We may be interested in the most disruptive mass gains, which occur when
the progenitor ratios are smallest.  We refer to these events as major
mergers.
Figure \ref{fig:merger1}, top, also shows the fraction of recently merged
clusters as a function of time for minimum mass ratios
1:3, 1:5 and 1:10.
For a lookback time of $7\,$Gyr ($z\sim 0.8$) for instance and the
$2.5\,$Gyr relaxation time, this means considering predecessors $9.5\,$Gyr
ago (at $z\sim 1.6$).  

The error bars are calculated using binomial statistics.  If $M$ of the
$N$ clusters have merged the most likely\footnote{The mean value is
$(M+1)/(N+2)$.} fraction is $f=M/N$ with variance: 
\begin{equation}
\sigma_f^2 = \frac{M(N-M)+1+N}{(N+2)^2(N+3)} \; .
\end{equation}
We use symmetric error bars.
One can see here that the number of clusters which have had a recent major
merger increases in the past, reaching a dramatic 80\% for 1:10 or smaller
mergers $7\,$Gyr ago (around $z \sim 0.83$) for a relaxation time of
$2.5\,$Gyr.
For shorter relaxation times fewer mergers have occurred, as expected.
For fixed relaxation times the
amount of accretion relative to merging changes with lookback time.

In the lower two panels different relaxation times are considered.
For some phenomena associated with mergers the relaxation times of interest
depend on the time of observation.
We saw in the previous section that departures from the virial relation
lasted several hundred Mpc of conformal time.  This is not unexpected.
A typical relaxation time is likely some fraction of the halo dynamical
time.  Since clusters have $\bar{\rho}\sim 10^2\rho_{\rm crit}$ and
characteristic times scale as $\rho^{-1/2}$, typical timescales should be
$0.1\,t_{\rm H}$ where $t_{\rm H}\equiv H^{-1}$ is the Hubble time.
Ref.~\cite{RowThoKay04} noted that X-ray disturbances in their gas
simulations lasted $\delta\tau=300\,h^{-1}$Mpc (scaling as $a$ rather
than the crossing time $\sim a^{3/2}$).
This corresponds to roughly $1.3\,h^{-1}$Gyr at the present, and shorter
times in the past.
Simulations reported in \cite{RicSar01} have major merger related disturbances
lasting approximately one crossing time.  From \cite{RosBorNor02}, we take
this to be $1\,$Gyr at the present.
These two relaxation times are used in Fig.~\ref{fig:merger1} bottom, with
that found by Ref.~\cite{RowThoKay04} at left and $a^{3/2}\,$Gyr at right.  
For the latter case, the $a$ used is for the initial time, before the
merger, as clusters which merge later have a larger $a$
and thus longer relaxation time, and thus will also still be unrelaxed
at the given lookback time.
The ratio between the fractions with mergers and large $\Delta M$ 
does not change as much with lookback time compared to the case where 
the relaxation time was independent of the lookback time.  This suggests
that equating large $\Delta M$ jumps with mergers is
more reliable for phenomena whose relaxation times scale with $a$.

\begin{figure*}[htb]
\begin{center}
\resizebox{5.5in}{!}{\includegraphics{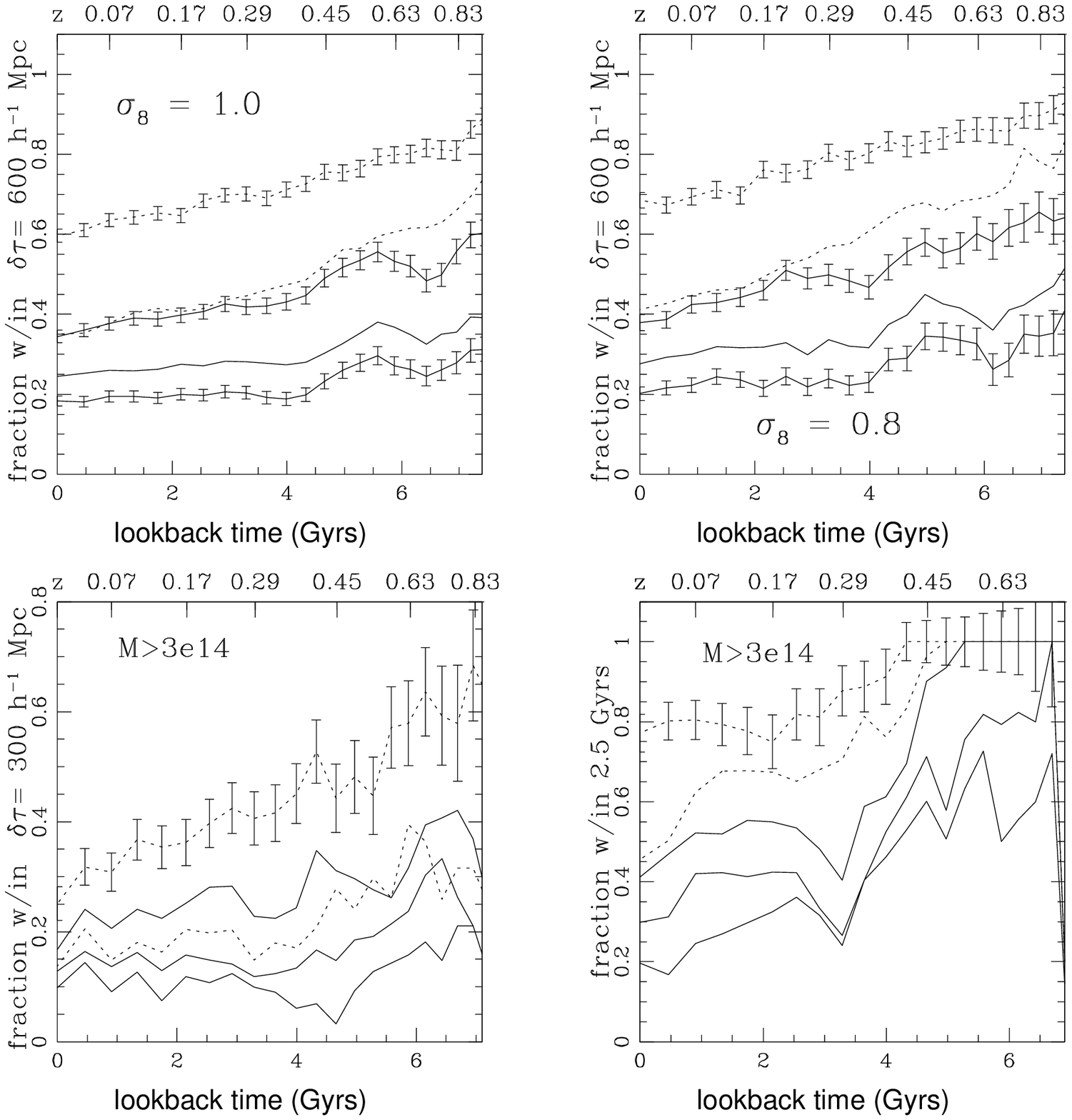}}
\end{center}
\caption{Top: recent mergers or large $\Delta  M$ for a relaxation time
of $\delta\tau = 600\,h^{-1}$Mpc, for two values of $\sigma_8$.
Bottom: the recently merged/large $\Delta M$ fraction for clusters with
$M>3.0 \times 10^{14}\,h^{-1}M_\odot$, for relaxation times of
$\delta\tau = 300\,h^{-1}$Mpc and $2.5\,$Gyr.
Smooth lines are 1:10, 1:5 and 1:3 predecessor mass ratios, dotted lines
are $M_f/M_i\geq 1.2, 1.33$ (top to bottom).  Errors as before.}
\label{fig:merger2}
\end{figure*}

In Fig.~\ref{fig:merger2} we compare different samples of clusters,
changing cosmology ($\sigma_8 \rightarrow 1.0$) and mass 
(taking $M>3.0 \times 10^{14}\,h^{-1}M_\odot$).
At the top are the fractions of recently merged or recent large $\Delta M$
clusters for both $\sigma_8 = 0.8$ and $\sigma_8 = 1.0$, for
the relaxation time is  $\delta\tau = 600\,h^{-1}$Mpc.
At present this lookback time is slightly over $2.5\,$Gyr.
The error bars are smaller for $\sigma_8=1.0$ because there are nearly
twice as many clusters in the sample.
The $\sigma_8=1.0$ clusters have had fewer mergers and mass jumps than
those for $\sigma_8=0.8$; clustering is less evolved for $\sigma_8=0.8$
so a cluster at fixed mass is more likely to be forming in our redshift
range.
At the bottom, the recently merged or recent large $\Delta M$ clusters
are shown for more massive clusters, $M >3.0\times 10^{14}\,h^{-1}M_\odot$,
with $\sigma_8 = 0.8$, for relaxation times of 
$\delta\tau = 300\,h^{-1}$Mpc (left) and for $2.5\,$Gyr (right).
As there are significantly fewer clusters (79 rather than 574) at $z=0$,
the error bars are a lot larger and are only shown for the top line.
Comparing with the same plots in Fig.~\ref{fig:merger1}, it is seen
that more of the massive clusters have had recent major mergers.

Values for $\sigma_8=1$ are reported in Tables \ref{tab:1.0Gyr}-\ref{tab:sound}
for four lookback times, along with $N_{\rm clus}$ so that the errors can
be estimated.

\begin{table*}
\begin{center}
\begin{tabular}
{c|c|c|c|c|c|c}
lookback time & 1:3 & 1:5 & 1:10 &20\% &25\% & $N_{\rm clus}$ \\ \hline 
0.0 &  0.07 &  0.08 &  0.12 &   0.14 &  0.08 & 909 \\  \hline 
2.2 &  0.08 &  0.11 &  0.18 &   0.23 &  0.13 & 739 \\  \hline 
4.0 &  0.09 &  0.13 &  0.22 &   0.34 &  0.15 & 573 \\  \hline 
6.2 &  0.19 &  0.26 &  0.39 &   0.60 &  0.35 & 346
\end{tabular}
\end{center}
\caption{The number of mergers or large $\Delta M$ events within the
previous $1\,$Gyr as a function of lookback time for our $\sigma_8=1.0$
simulation, comparable to Fig.~\ref{fig:merger1} upper left for
$\sigma_8 = 0.8$.}
\label{tab:1.0Gyr}
\end{table*}

\begin{table*}
\begin{center}
\begin{tabular}
{c|c|c|c|c|c|c}
lookback time & 1:3 & 1:5 & 1:10 &20\% &25\% & $N_{\rm clus}$ \\ \hline
0.0 &  0.18 &  0.24 &  0.34 &   0.59 &  0.34 & 909 \\  \hline 
2.2 &  0.22 &  0.31 &  0.45 &   0.73 &  0.49 & 739 \\  \hline 
4.0 &  0.29 &  0.38 &  0.57 &   0.83 &  0.67 & 573 \\  \hline 
6.2 &  0.45 &  0.54 &  0.72 &   0.97 &  0.90 & 346
\end{tabular}
\end{center}
\caption{The number of mergers or large $\Delta M$ events within the
previous $2.5\,$Gyr as a function of lookback time for our $\sigma_8=1.0$
simulation, comparable to Fig.~\ref{fig:merger1} upper right for
$\sigma_8 = 0.8$.}
\label{tab:2.5Gyr}
\end{table*}

\begin{table*}
\begin{center}
\begin{tabular}
{c|c|c|c|c|c|c}
lookback time & 1:3 & 1:5 & 1:10 &20\% &25\% & $N_{\rm clus}$ \\ \hline
0.0 &  0.08 &  0.11 &  0.16 &   0.23 &  0.12 & 909 \\  \hline 
2.2 &  0.10 &  0.13 &  0.21 &   0.28 &  0.16 & 739 \\  \hline 
4.0 &  0.09 &  0.12 &  0.21 &   0.33 &  0.15 & 573 \\  \hline 
6.2 &  0.16 &  0.22 &  0.32 &   0.48 &  0.26 & 346
\end{tabular}
\end{center}
\caption{The number of mergers or large $\Delta M$ events within the
previous $\delta\tau=300\,h^{-1}$Mpc as a function of lookback time
for our $\sigma_8=1.0$ simulation, comparable to Fig.~\ref{fig:merger1} 
lower left for $\sigma_8 = 0.8$.}
\label{tab:tau}
\end{table*}

\begin{table*}
\begin{center}
\begin{tabular}
{c|c|c|c|c|c|c}
lookback time & 1:3 & 1:5 & 1:10 & 20\% &25\% & $N_{\rm clus}$ \\  \hline
0.4 &  0.06 &  0.08 &  0.12 &   0.13 &  0.09 & 872 \\  \hline 
2.3 &  0.08 &  0.11 &  0.19 &   0.21 &  0.12 & 718 \\  \hline 
4.3 &  0.10 &  0.13 &  0.23 &   0.33 &  0.15 & 516 \\  \hline 
6.3 &  0.14 &  0.18 &  0.33 &   0.57 &  0.26 & 291
\end{tabular}
\end{center}
\caption{The number of mergers or large $\Delta M$ events within the
previous $a^{3/2}\,$Gyr as a function of lookback time for our $\sigma_8=1.0$
simulation, comparable to Fig.~\ref{fig:merger1} lower right for
$\sigma_8 = 0.8$.}
\label{tab:sound}
\end{table*}

\begin{figure*}[htb]
\begin{center}
\resizebox{5.5in}{!}{\includegraphics{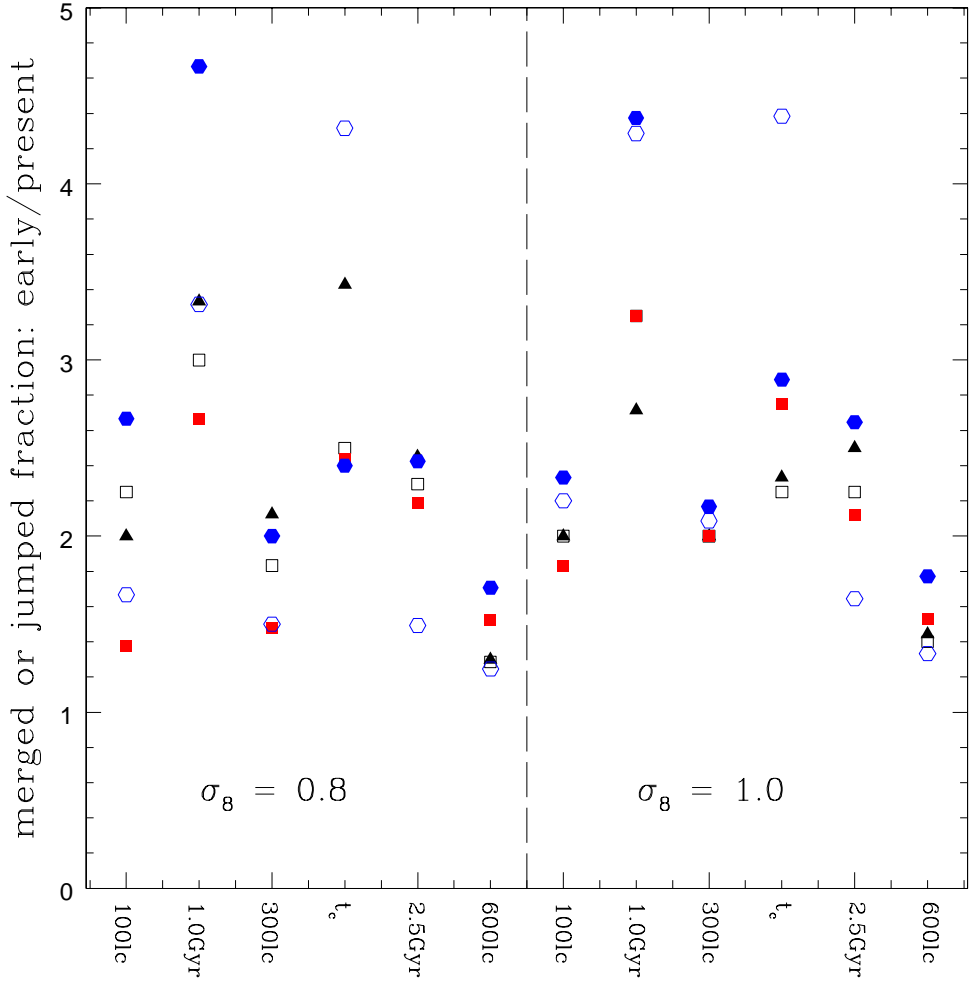}}
\end{center}
\caption{Summary of change in recent merger/recent large $\Delta M$ fraction
of clusters with time: (fraction $6.2\,$Gyr ago)/(fraction at present) for
most, (fraction $6.3\,$Gyr ago)/(fraction $0.4\,$Gyr ago) for the $a^{3/2}$
relaxation time case.  The various relaxation times are listed at the bottom:
100t, 300t, 600t ($\delta\tau=100, 300, 600\,h^{-1}$Mpc respectively),
$t_{\rm cross}=a^{3/2}\,$Gyr, and 1.0 and $2.5\,$Gyr.
Filled triangles are 1:3 mergers, open squares are 1:5 mergers, filled
squares are 1:10 mergers, open circles are 20\% mass jumps, filled circles
are 33\% mass jumps.  The errors are fairly large at the early time but one
can see that a doubling of the recently merged or large $\Delta M$ fraction
between the present and $z\sim 0.8$ is not uncommon and sometimes even
larger increases occur.}
\label{fig:merger3}
\end{figure*}

For all relaxation times there is an increase in the fraction of recently
merged or recent large $\Delta M$ as the lookback time increases,
i.e.~there were more major mergers and large mass jumps in the past.
The fraction depends on the relaxation time chosen.
An overview is provided in Fig.~\ref{fig:merger3} where we compare the
ratios of
merged and large $\Delta M$ samples near the present to that of about
$6\,$Gyr ago, for several different relaxation times in our two simulations.
A doubling of the recently merged or recent large $\Delta M$ fraction 
between the present and $z\sim 0.67$ is not uncommon, and sometimes
even larger increases occur.
The errors in these ratios for the relaxation time of
$\delta\tau = 100\,h^{-1}$Mpc are quite large.  For $\sigma_8=0.8(1.0)$, from
44(29)\% for the 1:3 mergers to 32(28)\% for $M_f/M_i \geq 1.2$.
For the other relaxation times the errors on the ratio are the largest for the
1:3 mergers (10-25\%) and smallest for the mass jumps with $M_f/M_i\geq 1.2$
(3-12\%).  The errors on the other cases fall in between.
More details can be obtained from the earlier plots and the Tables.

Another question of interest might be: in how many major mergers do three
or four predecessors have comparable mass?  
We took one particular definition as described in the methods section. 
These three-body mergers were fairly common, the exact numbers depend upon
relaxation times considered.  For the $\sigma_8 = 0.8$ case, about 15\% of
the 1:5 mergers were three-body for $\delta\tau=30\,h^{-1}$Mpc and slightly
less than 30\% for $\delta\tau=600\,h^{-1}$Mpc on average, with little
evidence of growth with time.
For $\sigma_8=1.0$ the fraction of three-body mergers to 1:5 major mergers
was closer to 10\% on average and seemed to grow with time for
$\delta\tau=300\,h^{-1}$Mpc.  For $\delta\tau=600\,h^{-1}$Mpc this growth
was more evident, starting around 12\% and growing to around 30\%.
For fixed relaxation times, 2.5 (1.0) Gyr, the fraction of three-body major
mergers to 1:5 major mergers showed a definite increase with lookback time.
For $\sigma_8=0.8$ it started at 30\% (10\%) and grew to 80\% (35\%) at
lookback times of $7\,$Gyr.
For $\sigma_8=1.0$ the fraction started at 12\% (5\%) and grew to
55\% (25\%) at lookback times of $7\,$Gyr.  
This means that even though a 1:5 merger might not be considered ``major''
by some definitions, it is very possible that there are two of these going
on at the same time, making these mergers more energetic. 

There were many fewer four-body mergers.  The largest effect was for the
$2.5\,$Gyr relaxation time, where four-body mergers were about one third
of the three-body mergers.  In the largest case, for $\sigma_8=0.8$ and
$2.5\,$Gyr relaxation time, the total fraction of clusters which had had
a recent four-body merger only reached 5\% or above for lookback times
greater than $4\,$Gyr.

\section{Conclusions} \label{sec:conclusions}

Clusters of galaxies represent the current endpoint of structure
formation.  As the largest systems which have had time to virialize
in a universe with hierarchical structure formation, they make excellent 
laboratories for cosmology, large-scale structure and galaxy formation.
The formation of galaxy clusters via a combination of mergers and 
accretion of smaller objects is crucial to understanding many of
their present day properties.

In this paper we investigated the formation of galaxy clusters in
some detail, and the major events which define this process,
extending earlier studies 
\cite{TorBouWhi97,Tor98,Col99,GotKlyKra01,Zha,RowThoKay04,TKGK,
Bus03,analytic} discussed in \S\ref{sec:background}.
While galaxy cluster formation properties can be reliably
calculated with the present generation of large, high-resolution N-body
simulations, it is difficult to find specific numbers in the literature.
To remedy this we have calculated, for a sample of hundreds of
clusters, the degree of virialization, formation time, and for two values of 
the clustering amplitude $\sigma_8$, the fraction of disturbed galaxy 
clusters for many definitions of disturbed.

We began by showing the time histories of a few clusters to illustrate
the typical formation pattern: periods of smooth accretion punctuated by 
large increases in mass.  We then turned to various characterizations
of this process, applied to our statistical sample as a whole.

We first calculated the ratio of kinetic to potential energy.
At early times clusters are hotter than the vacuum virial relation
(2KE$=$PE) would predict due to continuous infall of material.
With the onset of $\Lambda$ domination and the cessation of structure
growth the excess kinetic energy drops and their mass accretion rate slows.

We then turned to formation time definitions in the literature;
while no single number can capture the complexity of a cluster merging tree,
the concept of a formation time encodes much useful information.
We compared different formation time definitions which are relevant for
different types of observations.
For instance a recent large mass jump may not be of interest if one only
wants to know when the cluster first was detectable by 
SZ decrement (which presumably depends more on the depth of the potential 
well), but may be relevant for studies which
rely on a cluster being old enough to be dynamically relaxed.   
The different formation times we considered are correlated, with large 
scatter.  Formation times relying upon a smooth parameterized fit to
galaxy cluster histories were most correlated with each other, but the
results were sensitive to the fit methodology, due to the large amount
of recent sporadic mass gain.
We compared the distributions of the different formation times for our
whole cluster sample and for our more massive clusters, again finding some
correlation but also significant scatter.

We then turned from smooth parameterizations to the characteristic
abrupt jumps in mass over time for galaxy clusters.  On average, galaxy 
clusters have had at least 4 large mass jumps since $z \sim 2$, this number
increases with cluster mass; about half of the clusters
get at least half of their mass in these jump events.

The above measurements were for our $\sigma_8 = 0.8$ sample of 574 clusters.
The second part of the paper reported the time dependence of the
fraction of clusters which have had a recent mass jump
or major merger, using in addition a $\sigma_8 =1.0$ sample of 909 clusters.
These fractions should be
of use for estimates of the number of ``relaxed'' clusters available
in surveys (given a relaxation time for the phenomena of interest)
and for helping to constrain relaxation times for phenomena associated
with mergers or mass jumps where the phenomenon's occurrence fraction 
has been measured but not its relaxation time.
In Figs.~\ref{fig:merger1}-\ref{fig:merger3} and tables
\ref{tab:1.0Gyr}-\ref{tab:sound}
we give the merger fractions for a number of different situations.
These fractions also serve to quantify the previously seen trend of
more recent mergers (fractionally) at higher redshift.
Comparing $z=0$ with $z=0.67$ (about $6\,$Gyr ago), the fraction of clusters
which had a recent merger or mass jump is doubled by almost any definition
and for some definitions the increase is even larger.

The simulations used here were performed on the IBM-SP at NERSC.
JDC thanks T. Abel, S. Allen, G. Bryan, R. Gal, J. Hennawi, R. Kneissl, 
A. Kravtsov and
P. Ricker for helpful discussions and was supported in part by
NSF-AST-0205935. 
MJW was supported in part by NASA and the NSF.

\end{document}